\documentclass[prl,twocolumn,showpacs,groupedaddress]{revtex4}
\usepackage[dvips]{epsfig}
\usepackage{subfigure}
\usepackage{float}
\usepackage{amsmath}
\usepackage{amssymb}
\usepackage{amsfonts}
\usepackage{rotating}
\usepackage{euscript}
\usepackage{subfigure}
\usepackage{enumerate}
\usepackage{hhline}
\usepackage{threeparttable}
\usepackage{supertabular}
\usepackage{pslatex}
\usepackage{tabularx}
\usepackage{color}

\begin{document}

\title{\bf{Optical fiber taper coupling and high-resolution wavelength tuning of microdisk resonators at cryogenic
temperatures}}

\author{Kartik Srinivasan}
\email{phone: (626) 395-6269, fax: (626) 795-7258, e-mail:
kartik@caltech.edu}
\author{Oskar Painter}
\affiliation{Center for the Physics of Information and Department
of Applied Physics, California Institute of Technology, Pasadena,
CA 91125}

\date{\today}

\begin{abstract}

  A system for studying microcavity resonators at cryogenic temperatures ($\sim$ 10 K) through evanescent coupling via
  optical fiber taper waveguides is reported, and efficient fiber coupling to AlGaAs microdisk cavities with embedded
  quantum dots is demonstrated. As an immediate application of this tool, we study high-resolution tuning of microdisk
  cavities through nitrogen gas adsorption, as first discussed by Mosor, et al \cite{ref:Mosor}.  By proper regulation
  of the nitrogen gas flow and delivery of the gas to the sample surface, continuous tuning can be achieved with modest
  gas flows, and overall wavelength shifts as large as 4 nm are achieved.

\end{abstract}
\pacs{42.70.Qs, 42.55.Sa, 42.60.Da, 42.55.Px}
\maketitle

\noindent Solid-state systems involving a semiconductor
microcavity coupled to a semiconductor quantum dot (QD)
\cite{ref:Gerard_book_chapter} offer a promising implementation of
cavity quantum electrodynamics (cQED) \cite{ref:Kimble2} for
quantum information processing and computing applications. In
addition to demonstrations of vacuum Rabi splitting in the
emission spectrum of a QD-microcavity system
\cite{ref:Reithmaier,ref:Yoshie3,ref:Peter}, the quality factor
($Q$) of wavelength-scale III-V semiconductor microcavities has
recently exceeded 10$^5$ \cite{ref:Srinivasan9,ref:Herrmann},
paving the way for cQED experiments in which the coherent
QD-photon coupling rate can greatly exceed the system's
dissipative rates. Also important is the development of an
efficient interface through which the microscopic cavity field can
be accessed from macroscopic optics.  Silica optical fiber tapers,
initially used as effective input-output couplers for silica
microcavities \cite{ref:Knight,ref:Cai}, have recently been used
to couple to high refractive index microcavities
\cite{ref:Srinivasan7}, including AlGaAs microdisk cavities with
embedded QDs \cite{ref:Srinivasan9}.

Here, we report on the development of a system that extends our
previous work, done at room temperature and pressure, to the high
vacuum ($\sim10^{-6}$ Torr), cryogenic ($\sim10$ K) environments
in which QD-based cQED experiments are done
\cite{ref:Reithmaier,ref:Yoshie3,ref:Peter}.  This system is used
to interrogate wavelength-scale GaAs/AlGaAs microdisk cavities
containing self-assembled InAs QDs, with information such as the
cavity transmission and QD emission spectrum obtained. In the
second part of the paper, we build on the work of Refs.
\cite{ref:Mosor,ref:Strauf}, using nitrogen (N$_{2}$) gas
adsorption to tune the resonant wavelength of microcavities in a
cryogenic environment.  By proper regulation of the N$_{2}$ flow
and delivery of the gas near the sample surface, potential
difficulties discussed in Ref. \cite{ref:Mosor} are overcome, and
reproducible, high-resolution tuning is achieved. Furthermore, the
fiber taper coupling allows for detailed investigation of the
N$_{2}$ adsorption process.

\begin{figure}[t]
\centerline{\includegraphics[width=8.3cm]{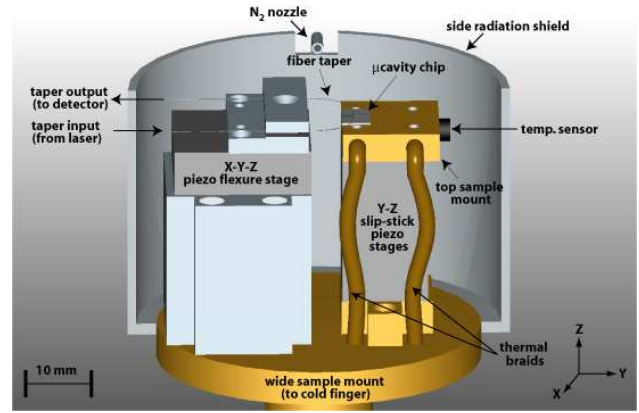}}
 \caption{Cut-away diagram of the cryostat interior,
 illustrating the arrangement of the fiber taper, sample, and N$_2$ flow nozzle.}
\label{fig:cryo_setup}
\end{figure}

The fiber tapers used are single mode optical fibers that have
been heated and stretched so that their central region has a
minimum diameter of $\sim1 \mu$m. Potential difficulties in
extending fiber taper coupling to a high vacuum, cryogenic
environment include the mechanical stability of the fiber taper
assembly, the lack of viscous air-damping of fiber taper
vibrations, and the mechanical and optical stability of the fiber
taper itself under repeated temperature cycling. Figure
\ref{fig:cryo_setup} depicts the setup we have developed, where
the sample and fiber taper reside inside a modified Janis ST-500
continuous flow, liquid He cryostat. A teflon-based compression
fitting \cite{ref:Abraham} is used to feed the two optical fiber
pigtails of the fiber taper from the interior vacuum to the
exterior of the cryostat. The fiber taper is held in a
``u''-shaped configuration to provide self-tensioning of the
taper. Coarse alignment of the taper to the microcavity is
achieved by positioning the microcavity sample using slip-stick
Y-Z piezo positioners with a displacement range of several
millimeters. Fine adjustment in the taper position is provided by
an X-Y-Z piezo-electric flexure stage with a maximum displacement
of several microns at $4.2$ K.  A thermally conductive pathway
between the sample and the cold finger is provided by copper
braids that connect the top sample mount to the cold finger. This
ensures that the sample can get to the requisite low temperature
($< 15$ K), which is measured by a silicon diode that is affixed
to the top sample mount.

\begin{figure}[t]
\centerline{\includegraphics[width=8.3cm]{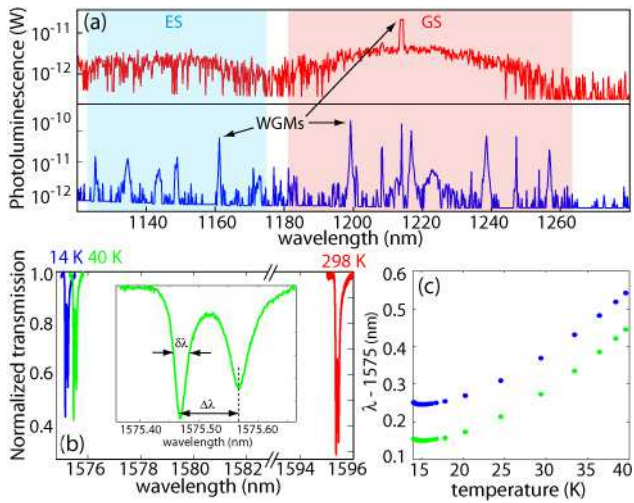}}
 \caption{(a) Comparison of photoluminescence data from an embedded layer of QDs in a
 $D$=4.5 $\mu$m disk, using free-space collection (top) and
fiber-taper-based collection (bottom). Resolution bandwidth is 0.1 (1)
nm for the fiber taper (free-space) collection spectra. (b)-(c) Temperature
tuning data for the TE$_{1,20}$ WGM.} \label{fig:PL_plus_temp_tuning}
\end{figure}

The devices we study are small diameter ($D$=2-4.5 $\mu$m)
GaAs/AlGaAs microdisks that contain a single layer of InAs QDs
(room temperature ground state emission at $\sim1317$ nm), as
discussed elsewhere \cite{ref:Srinivasan9}.  The cryostat is
cooled to a sample temperature of $14$ K, and during this process,
no additional loss in the optical fiber taper transmission is
observed (typical total insertion loss is 10-50\% depending upon
the taper tension). The taper is positioned in the near-field of
the microdisk under study using the piezo stage configuration
described above.  An adjustable air-gap may be maintained between
taper and disk or the taper may be placed into direct contact with
the disk, depending upon the level of cavity loading desired
(anywhere from under- to over-coupled is possible).  Accuracy in
the taper-disk gap is limited only by vibration-induced
fluctuations in the taper position (tens of nm in our current
set-up). In general, we have found that the taper can remain in a
fixed coupling position (at room or cryogenic temperatures) for
times as long as several hours.  The fiber taper was also
unaffected by repeated temperature cycling. As described in detail
in Ref.  \cite{ref:Srinivasan11}, the taper can be used to enhance
the collection efficiency of light emitted from microdisk
whispering gallery modes (WGMs) by nearly two orders of magnitude
over normal-incidence free-space collection.  Figure
\ref{fig:PL_plus_temp_tuning} compares fiber taper and free-space
collection of the low temperature (T$=14$ K) emission from a
$D=4.5\mu$m microdisk that is optically pumped with $\sim$100
$\mu$W of incident power from an 830 nm laser diode.  The
collected power and number of cavity modes observed in the
taper-collected spectrum greatly exceeds that obtained by
free-space collection, with emission into WGMs from the ground
(GS) and excited state (ES) manifolds of the QDs clearly visible.

More than just an efficient collection optic, the fiber taper may
also be used to optically probe and excite the cavity-QD system in
a highly efficient manner.  Here we use the fiber taper to monitor
the transmission properties of the cavity modes of the microdisk
as a function of temperature, and as described below, during
cavity mode tuning experiments involving N$_{2}$ gas adsorption.
To this end, a scanning tunable laser (linewidth $<5$ MHz) is
connected to the fiber taper input and the wavelength-dependent
taper output transmission is monitored with a photodetector. The
polarization state of the fiber taper mode is achieved through a
polarization controller inserted between the laser and taper
input.  Figure \ref{fig:PL_plus_temp_tuning}(b) shows the
transmission spectra of a cavity mode in a $D=4.5 \mu$m disk in
the 1500 nm band\footnote{From finite-element-method simulations,
this mode is
  identified as a TE (electric field components predominantly in-plane), $p=1$, $m=20$ WGM, where $p$ and $m$ denote the
  radial and azimuthal mode number, respectively.}. The cavity mode wavelength is seen to shift approximately $20$ nm as
the temperature is reduced from $298$ K to $14$ K as a result of the decrease in the AlGaAs disk refractive index. As
noted in our previous work (Ref. \cite{ref:Srinivasan9} and references therein), the high-$Q$ modes often appear as a
doublet (inset of Fig.  \ref{fig:PL_plus_temp_tuning}(b)) due to surface roughness on the microdisk which couples the
initially degenerate WGMs into frequency-split standing wave modes.

As discussed in many other works, the small tuning range (0.3 nm
for T$=14$-$40$ K in Fig. \ref{fig:PL_plus_temp_tuning}(c))
afforded by thermal tuning is a significant limitation in cQED
experiments due to the imprecise spectral positioning of QD
exciton peaks and cavity modes during growth and fabrication. To
overcome this difficulty, Mosor, et al., utilized noble gas
adsorption on the sample surface to achieve post-fabrication
shifts in a photonic crystal cavity of up to 5 nm\cite{ref:Mosor}.
Here, we apply this method to tune the resonances of our microdisk
cavities while using the fiber taper probe to monitor their
behavior.  Reference \cite{ref:Mosor} achieves wavelength tuning
in discrete steps by filling a secondary chamber with gas (Xe or
N$_{2}$) until a desired pressure is reached, releasing that
volume into the cryostat, and then repeating.  The authors found
that the fill pressure must lie within a vary narrow range, below
which no tuning occurs and above which excessively fast tuning
occurs.  To improve upon the tuning resolution and repeatability,
we have made two key modifications. Rather than introduce the gas
through the cryostat vacuum line, we inject it through a 1/16"
tube (inner diameter $0.56$ mm) that is routed into an opening in
the top of the side radiation shield (Fig. \ref{fig:cryo_setup}),
so that gas can be locally delivered with line-of-sight to the
sample.  Next, instead of introducing the gas into the cryostat
through repeated cycles, we fill an external chamber (V=0.1 L)
until a fixed pressure is reached (10 torr) and then release it
into the cryostat using a metering valve to control the flow rate.
We monitor the cavity mode transmission spectrum and use a
shut-off valve to stop the gas flow when a desired wavelength
shift is achieved (the shut-off and metering valves are positioned
as close as possible to the cryostat to minimize dead volume
between themselves and the end of the injection nozzle).

\begin{figure}
\centerline{\includegraphics[width=8.3cm]{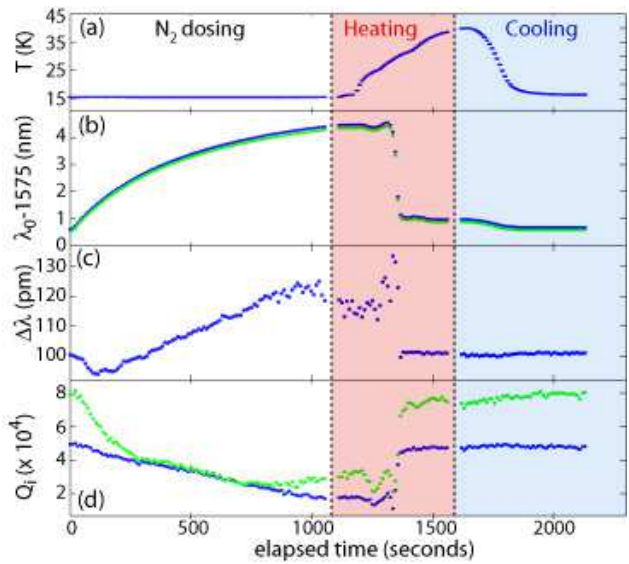}}
 \caption{Tuning data (TE$_{1,20}$ mode in a $D=4.5 \mu$m disk) during N$_{2}$ dosing,
 sample heating, and sample cooldown phases.  (a) Sample temperature (T), (b) resonant mode
 wavelengths ($\lambda_{0}$), (c) doublet mode splitting ($\Delta\lambda$),
 and (d) intrinsic cavity $Q$ ($Q_{i}$). The green (blue) curve is for the short (long)
wavelength doublet mode.}
 \label{fig:N2_tuning_data_first_flow}
\end{figure}

Tuning data obtained using high purity ($99.9995\%$) N$_{2}$ gas
is shown in Fig. \ref{fig:N2_tuning_data_first_flow} for the
microdisk mode studied in Fig. \ref{fig:PL_plus_temp_tuning}(b-c),
with the taper in contact with the disk edge.  Transmission
spectra of the taper-coupled microdisk were recorded every 10
seconds over the entire tuning cycle, which included an initial
N$_{2}$ dosing period, a subsequent N$_{2}$ desorption phase
accomplished by heating the sample with a resistive heater, and a
final cooldown period.  The cavity transmission spectra were fit
using a standard doublet model\cite{ref:Borselli2}, from which we
obtain the spectral position of the resonant modes
($\lambda_{0}$), the doublet mode splitting ($\Delta\lambda$), and
the intrinsic cavity $Q$-factor ($Q_{i}$). Figure
\ref{fig:N2_tuning_data_first_flow}(b) shows smooth, continuous
tuning is achieved, with a resonance wavelength shift of
$\Delta\lambda_{0}=3.8$ nm occurring after 1060 s of N$_{2}$
dosing.  During the heating phase the N$_{2}$ shut-off valve is
closed, and the wavelength dramatically drops at T$=28$ K as
N$_{2}$ begins to rapidly desorb from the disk
surface\cite{ref:Schlichting1}. The temperature is further raised
to $40$ K to ensure complete N$_{2}$ removal.  Finally, the sample
is cooled back down to the starting temperature, at which point
$\lambda_{0}$, $\Delta\lambda$, and $Q_{i}$ have returned to their
original values.

The tuning cycle of Fig.  \ref{fig:N2_tuning_data_first_flow} was
found to be repeatable from run-to-run, and could be interrupted
during the N$_{2}$ dosing phase to position the cavity mode
resonance wavelength with an accuracy of better than $\pm 10$ pm.
Once positioned, for temperatures below T$=20$K (where N$_{2}$
desorption is negligible over a timescale of
hours\cite{ref:Schlichting1}) we found the cavity mode wavelength
to be highly stable.  One non-ideal side effect of the N$_{2}$
tuning, evident in Fig.  \ref{fig:N2_tuning_data_first_flow}(d),
is the degradation in the $Q$-factor with increasing N$_{2}$
adsorption ($Q$-degradation factors of 2-3 for $4$ nm of tuning
were typical for modes of $Q\sim10^5$).  Several features in the
data of Fig.  \ref{fig:N2_tuning_data_first_flow} indicate that
the optical loss is due to sub-wavelength optical scattering from
the adsorbed N$_{2}$ film.  The clearest indicator is the rapid
rise in doublet splitting with wavelength tuning (Fig.
\ref{fig:N2_tuning_data_first_flow}(c)), a result of increased
surface scattering\cite{ref:Borselli3}.  Visual inspection of
microdisks after large tuning excursions also showed clouding of
the top surface.  Both observations point to an incomplete wetting
of N$_{2}$ on AlGaAs, and the growth of a rough bulk overlayer
consisting of N$_{2}$ crystallites\cite{ref:Volkmann1}.

\begin{figure}[t]
\centerline{\includegraphics[width=8.3cm]{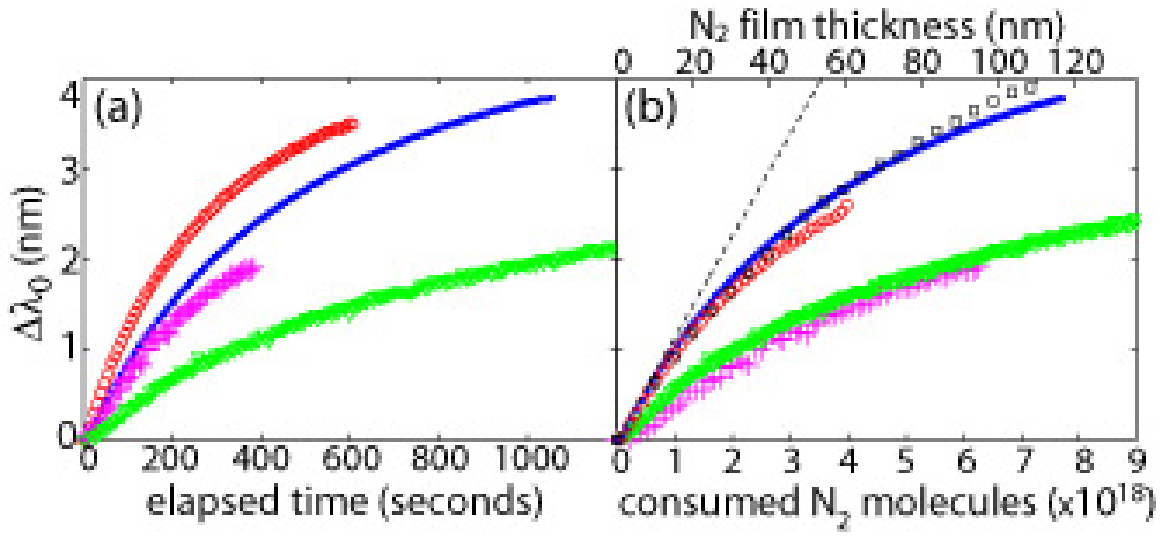}}
 \caption{Wavelength shift ($\Delta\lambda_{0}$) versus (a) time and (b) N$_{2}$ consumption
 (thickness) for different N$_{2}$ flow rates.  ($\textcolor{red}{\circ},\textcolor{blue}{\cdot}$)
 correspond to measurements of the mode in Fig. \ref{fig:N2_tuning_data_first_flow} ($D$=4.5 $\mu$m),
 while ($\textcolor{magenta}{+},\textcolor{green}{\triangledown}$) are for a WGM at $\lambda_{0}=1255$ nm in a
$D$=2.0 $\mu$m microdisk.  The \emph{average} flow rates are:
$\{\textcolor{red}{\circ},\textcolor{blue}{\cdot},
\textcolor{magenta}{+},\textcolor{green}{\triangledown}\}=\{13.0,7.4,16.0,5.5\}\times10^{15}$
N$_{2}$ molecules/s. The $\textcolor{black}{\square}$
($\textcolor{black}{--}$) curve in (b) corresponds to FEM
(perturbation) simulation of the tuning versus film thickness for
the $D$=4.5 $\mu$m TE$_{1,20}$ WGM.}
\label{fig:tuning_data_many_curves}
\end{figure}

Fig. \ref{fig:tuning_data_many_curves}(a) shows the wavelength
shift versus elapsed time, under varying flow conditions, and for
cavity modes of two different microdisks ($D=4.5,2.0$ $\mu$m). A
simple perturbative analysis relates the cavity mode tuning to the
mode's overlap with the disk surface through the equation
$\Delta\lambda_{0}\sim\lambda_{0}(n_{f}^{2}-1)\Gamma^{\prime}t_{f}/2$,
where $t_{f}$ is the film thickness, $n_{f}$ is the film
refractive index ($=1.21$\cite{ref:Pilla1}), and $\Gamma^{\prime}$
is the (linear) modal energy density at the air-disk interface.
Such an analysis yields an N$_{2}$ film thickness of
$t_{f}\gtrsim50$ nm for the measured mode tunings of $2-4$ nm. For
films of this thickness the perturbative analysis breaks down and
one must resort to more exact numerical methods. Finite-element
method (FEM) simulations of the mode tuning versus N$_{2}$ film
thickness were performed for the TE$_{1,20}$ mode of the $D=4.5$
$\mu$m microdisk, and are plotted in Fig.
\ref{fig:tuning_data_many_curves}(b) for N$_{2}$ coverage of the
top and side of the disk (assuming line-of-sight deposition).  The
measured data are also plotted in Fig.
\ref{fig:tuning_data_many_curves}(b) versus consumed N$_{2}$
(estimated from the initial and final chamber pressures, and
assuming an exponential decrease in the pressure with time).  The
simulated N$_{2}$ thickness is related to the measured N$_{2}$
consumption with a fixed scaling factor (the sticking coefficient
of rare gases to their solids is known to be
near-unity\cite{ref:Schlichting2}, i.e., constant), determined by
a least-squares fit to the measured data.  From these plots the
measured tuning is seen to be independent of flow rate for each of
the cavity modes, and the saturation in the tuning rate with
increasing layer thickness is well captured by the FEM simulation.

In summary, we have demonstrated that optical fiber tapers can provide an efficient interface for transferring light to
and from standard laboratory fiber optics into a micron-scale cavity housed in a high-vacuum, cryogenic environment.  In
addition, we have shown that rare gas adsorption can be used to produce high
resolution, continuous tuning of microdisk WGM wavelengths.  These two tools are of significant utility to future cQED
experiments involving interactions of single QDs with fiber-coupled microdisk cavities \cite{ref:Srinivasan13}.


The authors thank A. Stintz and S. Krishna of the University of
New Mexico for materials growth, and R. Heron of the
Janis Research Company for assistance in the cryostat design.

\bibliography{./PBG_10_29_2006}

\end{document}